\documentclass[aps,prd,twocolumn,superscriptaddress,nofootinbib,floatfix]{revtex4-2}
\usepackage{float}
\IfFileExists{newtxtext.sty}{%
  \usepackage[T1]{fontenc}%
  \usepackage{newtxtext}%
  \usepackage{newtxmath}%
}{%
  \usepackage{mathptmx}%
}

\usepackage{amsmath,amssymb,bm}
\usepackage{graphicx}
\usepackage{booktabs}
\PassOptionsToPackage{hyphens}{url}
\usepackage[colorlinks=true,linkcolor=blue,citecolor=magenta,urlcolor=blue]{hyperref}
\AtBeginDocument{\let\hbar\hslash}

\def\doi#1{\href{https://doi.org/#1}{\color{blue}#1}}
\usepackage{tikz}
\usetikzlibrary{shapes.geometric,shapes.symbols}
\DeclareRobustCommand{\orcidicon}{%
  \begin{tikzpicture}
    \definecolor{orcidgreen}{HTML}{166B3A}
    \draw[orcidgreen, fill=orcidgreen] (0,0)
    circle [radius=0.16]
    node[white] {{\fontfamily{qag}\selectfont \tiny ID}};
  \end{tikzpicture}
  \hspace{-2mm}
}
\foreach \x in {A, ..., Z}{%
  \expandafter\xdef\csname orcid\x\endcsname{\noexpand\href{https://orcid.org/\csname orcidauthor\x\endcsname}{\noexpand\orcidicon}}
}

\newcommand{\MPl}{M_{\mathrm{Pl}}}
\newcommand{\rhoL}{\rho_{\mathrm{v}}}
\newcommand{\rhoc}{\rho_{\mathrm{c}}}
\newcommand{\mm}{m_{\mathrm{m}}}
\newcommand{\pr}{p_{r}}
\newcommand{\pt}{p_{t}}
\newcommand{\peff}{p_{r,\mathrm{eff}}}
\newcommand{\rhoeff}{\rho_{\mathrm{eff}}}
\newcommand{\pteff}{p_{t,\mathrm{eff}}}
\newcommand{\creff}{c^{2}_{r,\mathrm{eff}}}
\newcommand{\cteff}{c^{2}_{t,\mathrm{eff}}}

\definecolor{ILdarkred}{rgb}{0.60,0.00,0.00}
\definecolor{ILdarkblue}{rgb}{0.00,0.00,0.55}
\definecolor{ILdarkgreen}{rgb}{0.00,0.38,0.00}
\newcommand{\ilnew}[1]{\relax\ifmmode{\color{ILdarkgreen}#1}\else
  {\color{ILdarkgreen}\bfseries #1}\fi}

\newcommand{\iltext}[1]{\relax\ifmmode{\color{ILdarkred}#1}\else
  {\color{ILdarkred}\bfseries #1}\fi}
\newcommand{\ilnum}[2]{\relax\ifmmode{\color{ILdarkred}[\mathrm{SUGGESTED:#1}]}\else
  {\color{ILdarkred}\bfseries [SUGGESTED VALUE, #1: #2]}\fi}

\newcommand{\ilfig}[3]{%
  \begin{figure}[H]
  \centering
  \includegraphics[width=1\columnwidth]{#1.png}  % [IL] figure now included
  \caption{#2}
  \label{fig:#1}
  \end{figure}}

\newcommand{\ilfigwide}[3]{%
  \begin{figure*}[htbp]
    \centering
    \includegraphics[width=0.95\textwidth]{#1.png}  % <--- Added this line!
    \caption{#2}
    \label{fig:#1}
  \end{figure*}
}

\begin{document}

\title{Anisotropic Dark Matter Stars with a Pressure-Sourced Dark Energy Density}

\author{Qazi Abdul Ghafoor\,\orcidB{}}
\affiliation{\href{https://ror.org/04c4dkn09}{Department of Astronomy, University of Science and Technology of China--USTC}, \\ Hefei, Anhui 230026, People's Republic of China}
\affiliation{\href{https://ror.org/04c4dkn09}{School of Astronomy and Space Science, University of Science and Technology of China--USTC}, \\ Hefei, Anhui 230026, People's Republic of China}

\author{Il\'idio Lopes\,\orcidA{}}
\affiliation{\href{https://ror.org/020hp3377}{Centro de Astrof\'isica e Gravita\c{c}\~ao--CENTRA}, Departamento de F\'isica, \\ Instituto Superior T\'ecnico--IST, \href{https://ror.org/01c27hj86}{Universidade de Lisboa--UL}, \\ Av.~Rovisco Pais~1, 1049-001 Lisboa, Portugal}

\date{\today}

\begin{abstract}
If the dark sector is not inert, the vacuum energy trapped inside a dark
matter star need not equal its cosmological value, but may instead respond
to the star's own density and pressure. We show that this freedom is far
more restricted than it first appears: demanding a smooth, physically
sensible stellar surface rules out the response one would most naturally try,
in which the vacuum energy simply tracks the local density, since it drives
the sound speed superluminal in the outer layers. The simplest response that
survives is one sourced by the local pressure rather than the density, and we
adopt it within an established anisotropic dark matter star model, obtaining
the full stellar structure in closed form. The resulting star is heavier than
its uncoupled counterpart, by almost $30\%$ at the causal limit, yet
correspondingly less strongly supported by pressure, so that the coupling is
bounded not by any failure of pressure support but by the star's compactness
approaching the threshold for gravitational collapse. We show, further, that
once this response is switched on, causality is always the tightest physical
requirement upon the star, rendering the dominant energy condition redundant
throughout the admissible parameter space.
\end{abstract}

\maketitle

\section{Introduction}
\label{sec:intro}

Dark matter constitutes the greater part of the matter content of the
Universe, and yet almost everything that is known of it has been inferred from
its gravitational influence upon luminous tracers. Should the dark sector
possess self-interactions of sufficient strength, dark matter may cool, lose
energy and collapse into gravitationally bound configurations of its own. The
resulting objects, commonly termed dark matter stars, offer a rare prospect of
studying the dark sector through the structure of a compact object rather than
through its ambient distribution.

In the absence of a settled microphysics it has become customary to adopt an
effective polytropic equation of state and to permit the pressure to be
anisotropic. Anisotropy of this kind is neither exotic nor introduced for
convenience; it arises generically in the presence of strong self-interactions,
of solid or superfluid phases, and of magnetic fields, and its consequences for
relativistic spheres have been studied since the work of Bowers and
Liang~\citep{BowersLiang1974} and reviewed by Herrera and
Santos~\citep{HerreraSantos1997}. Moraes, Panotopoulos and
Lopes~\citep{Moraes2021} constructed anisotropic dark matter stars on these
foundations and obtained configurations whose masses and radii fall within the
range now probed by pulsar timing and by gravitational-wave observation.
A fundamental question in the study of such objects concerns the maximum possible compactness achievable within General Relativity. Buchdahl (1959) addressed this first by showing  that the surface compactness is bounded by \(2M/R \leq 8/9\) for a spherically symmetric perfect fluid with non-negative non-increasing density and finite central pressure \cite{Buchdahl1959, buchdahl1966}. This result establishes a crucial restriction: a fluid configuration cannot attain compactness values arbitrarily close to the black hole limit within general relativity, for which \(2M/R = 1\) \cite{cardoso2019}. The Buchdahl limit therefore serves as a fundamental benchmark for distinguishing between stable compact stars and black holes, and for investigating the nature of compact relativistic objects \cite{cardoso2019}.

That treatment, in common with most work on relativistic stellar structure,
sets the cosmological constant to zero. Static spheres with a constant vacuum
energy density have nonetheless been studied for some time, by
B\"ohmer~\citep{Boehmer2004}, by B\"ohmer and Harko~\citep{BoehmerHarko2005}
and by Gabbanelli and collaborators~\citep{Ovalle2019}. In every case the vacuum
energy density is a universal constant, and the corrections to stellar
structure are correspondingly minute: for the observed value they are smaller
than the terms they correct by some forty-three orders of
magnitude~\citep{Weinberg1989}.

The present work takes a different route. If dark matter and dark energy
interact~\citep{Wang2016, Amendola2000}, the vacuum energy density is not universal
but responds to its environment, and inside a compact object it may be enhanced by many orders of
magnitude without any conflict with cosmology. A vacuum energy density confined
to the interior of a dense region is in any case familiar: the bag constant of
the MIT model~\citep{Chodos1974,FarhiJaffe1984}, which fixes the density scale
of Ref.~\citep{Moraes2021} through $\rhoc = 14B$ and $K = z/B$, with $z$
the dimensionless self-interaction parameter of that model, enters the
energy-momentum tensor as $T_{\mu\nu} = -Bg_{\mu\nu}$, precisely the form of a
cosmological term.

The question is then which local responses are admissible. We show in
Sec.~\ref{sec:surface} that the stellar surface answers it almost completely,
and that the answer excludes the most obvious choice. Our principal result is
accordingly a constraint rather than a model: the conditions obtained there
restrict any local response whatever, and the particular response adopted in
the remainder of the paper is simply the first power law that satisfies them. Section~\ref{sec:model}
fixes the conventions and the matter content, Sec.~\ref{sec:structure} obtains
the structure equations, Sec.~\ref{sec:conditions} establishes the admissible
region of parameter space, Sec.~\ref{sec:results} presents the consequences,
and Sec.~\ref{sec:conclusion} discusses what the model does and does not imply.

\section{The surface conditions on a local response}
\label{sec:surface}

Let the vacuum energy density be a local function of the dark matter density,
\begin{equation}
\rhoL(r) = f\bigl(\rho(r)\bigr), \qquad f \geq 0, \quad f' \geq 0 ,
\label{eq:response}
\end{equation}
so that the vacuum contribution is non-negative and grows with the matter
density. Transposing $T^{\mu\nu}_{\rm v} = -\rhoL g^{\mu\nu}$ to the
right-hand side of the field equations gives an effective fluid with
\begin{equation}
\rhoeff = \rho + f, \qquad \peff = \pr - f, \qquad
\pteff = \pt - f .
\label{eq:effective}
\end{equation}

Two requirements at the zero-density surface constrain $f$ severely. The first
is that the effective radial pressure vanish there, so that the exterior is
vacuum and the Israel--Darmois conditions~\citep{Israel1966} are satisfied
without a surface layer or a residual anisotropy. Since $\pr(R) = 0$, this
requires
\begin{equation}
f(0) = 0 .
\label{eq:cond1}
\end{equation}
The second is that the radial sound speed remain non-negative as the surface is
approached. From Eq.~(\ref{eq:effective}),
\begin{equation}
\creff = \frac{d\peff}{d\rhoeff}
= \frac{\dfrac{d\pr}{d\rho} - f'(\rho)}{1 + f'(\rho)} ,
\label{eq:cs}
\end{equation}
and for an equation of state with $d\pr/d\rho \to 0$ as $\rho \to 0$, as holds
for any polytrope of index greater than unity, the surface value is
$-f'(0)/[1+f'(0)]$. Non-negativity therefore requires
\begin{equation}
f'(0) = 0 .
\label{eq:cond2}
\end{equation}

Equations~(\ref{eq:cond1}) and~(\ref{eq:cond2}) dispose at once of the linear
response $f = \beta\rho$, which satisfies the first but not the second and
yields $\creff(R) = -\beta/(1+\beta) < 0$: an imaginary sound speed
throughout the outer layers, and a boundary pathology rather than a physical
prediction. Among power laws $f \propto \rho^n$ the conditions require
$n > 1$, and the first admissible integer choice is $n = 2$. With the
polytropic equation of state adopted below this is the pressure-sourced
response
\begin{equation}
\rhoL = \alpha K \rho^2 = \alpha\,\pr, \qquad 0 \leq \alpha < 1 .
\label{eq:ansatz}
\end{equation}
The upper limit is required because $\alpha = 1$ annihilates the effective
radial pressure identically. We adopt Eq.~(\ref{eq:ansatz}) for the remainder
of this work, noting that it has a natural reading: the vacuum responds not to
the amount of dark matter present but to the pressure it exerts.

\section{The model}
\label{sec:model}

Throughout we adopt the signature $(-,+,+,+)$ and Planck units, in which
$\hbar = c = G = 1$ and therefore $\MPl = 1$; densities and pressures are then
measured in $\MPl^4$ and lengths in Planck lengths. No factor of $\hbar$ or $c$
appears anywhere below. We consider
static, spherically symmetric configurations,
\begin{equation}
ds^2 = -e^{\nu(r)}dt^2 + \left(1 - \frac{2m(r)}{r}\right)^{-1}dr^2
+ r^2 d\Omega^2 ,
\label{eq:metric}
\end{equation}
where $m(r)$ is the total Misner--Sharp mass, comprising matter and vacuum
alike; we reserve $\mm(r)$ for the matter mass.

Following Ref.~\citep{Moraes2021} the dark matter is described by an
anisotropic energy-momentum tensor with
\begin{equation}
\pr = K\rho^2, \qquad
\rho(r) = \rhoc\left(1 - \frac{r^2}{R^2}\right),
\label{eq:eos}
\end{equation}
and $K = z/B$, $\rhoc = 14B$, so that the single dimensionless
combination $K\rhoc = 14z$ governs everything that follows. With
Eq.~(\ref{eq:ansatz}) the effective fluid of
Eq.~(\ref{eq:effective}) becomes
\begin{equation}
\rhoeff = \rho\,(1 + \alpha K\rho), \qquad
\peff = (1-\alpha)\,K\rho^2 ,
\label{eq:eff_explicit}
\end{equation}
and the inertial combination is again untouched,
\begin{equation}
\rhoeff + \peff = \rho + \pr ,
\label{eq:inertial}
\end{equation}
so that the response alters the pressure that resists compression, and the
gravitating energy, but not the inertia of the fluid.

Because $\rhoL$ is not constant the two sectors are not separately conserved.
Total conservation gives
\begin{equation}
\nabla_\mu T^{\mu\nu}_{\rm dm} = \partial^\nu \rhoL
= \alpha\,\partial^\nu \pr ,
\label{eq:exchange}
\end{equation}
a single exchange term proportional to the pressure gradient.

\section{Structure equations}
\label{sec:structure}

The temporal field equation gives the mass function by quadrature. With
Eqs.~(\ref{eq:eos}) and~(\ref{eq:eff_explicit}) the integral is elementary.
The matter mass is
\begin{equation}
\mm(r) = 4\pi\rhoc r^3\left(\frac{1}{3} - \frac{r^2}{5R^2}\right),
\qquad \mm(R) = \frac{8\pi}{15}\rhoc R^3,
\label{eq:mm_r}
\end{equation}
and the vacuum contribution, from \(\rhoL = \alpha K\rho^2\), is
\begin{equation}
m_{\rm v}(r) = 4\pi\alpha K\rhoc^2 r^3\left(\frac{1}{3} - \frac{2r^2}{5R^2} + \frac{r^4}{7R^4}\right).
\label{eq:mv_r}
\end{equation}
The total mass is therefore
\begin{equation}
m(r) = \mm(r) + m_{\rm v}(r),
\label{eq:m_full}
\end{equation}
and at the surface,
\begin{equation}
\frac{M}{\mm(R)} = 1 + \frac{4}{7}\,\alpha K\rhoc .
\label{eq:mass}
\end{equation}
Since $K\rhoc = 14z$, this is simply
\begin{equation}
\frac{M}{\mm(R)} = 1 + 8\,\alpha z ,
\label{eq:mass_z}
\end{equation}
so the enhancement is at most $7.2\%$ at $z = 0.009$, and $28.6\%$ at the
causal limit $z = 1/28$.
Unlike a constant vacuum term, whose contribution to the Misner--Sharp mass is
not inferred by an exterior observer, the vacuum here is bound to the star and
vanishes outside it, so Eq.~(\ref{eq:mass}) is a genuine enhancement of the
gravitating mass and of the compactness.

Hydrostatic equilibrium follows from the conservation of the total
energy-momentum tensor. Reducing the anisotropic Tolman--Oppenheimer--Volkoff
equation for the effective fluid by means of Eqs.~(\ref{eq:eff_explicit})
and~(\ref{eq:inertial}), as set out in Appendix~\ref{app:structure}, gives
\begin{equation}
\pr' = -\frac{(\rho + \pr)[m + 4\pi r^3 (1-\alpha)\pr]}
{(1-\alpha) r^2(1 - 2m/r)}
+ \frac{2\Delta}{(1-\alpha) r}.
\label{eq:tov_final}
\end{equation}
where $\Delta \equiv \pt - \pr$ is the anisotropic factor. Because the
vacuum is isotropic it cancels from the difference of the principal pressures,
so that $\Delta_{\rm eff} \equiv \pteff - \peff = \Delta$.
For \(\alpha = 0\), Eq.~(\ref{eq:tov_final}) reduces to the anisotropic
TOV equation of Ref.~\citep{Moraes2021}. As \(\alpha \to 1\) the
effective radial pressure vanishes identically and Eq.~(\ref{eq:tov_final})
degenerates, radial support being lost; that limit is excluded here rather than
treated.

The metric potential, the Israel--Darmois matching and the closed form of
the anisotropy are obtained in the same way and are collected in
Appendix~\ref{app:structure}. Two of their consequences are used below. The
exterior is Schwarzschild and the matching requires no thin shell, since $\rho$,
$\pr$ and $\rhoL$ all vanish at $r = R$. And the anisotropy vanishes at both
boundaries, $\Delta(0) = \Delta(R) = 0$, at the centre by regularity and at the
surface because $\rho$, $\pr$ and $\pr'$ do; its sign in the interior depends
upon $\alpha$, and for $\alpha = 0$ it is positive throughout, recovering
Ref.~\citep{Moraes2021}.

The tangential pressure is
\begin{equation}
\pt = \pr + \Delta,
\label{eq:pt_def}
\end{equation}
and the effective tangential pressure is then
\begin{equation}
\pteff = (1-\alpha)\pr + \Delta.
\label{eq:pt_eff}
\end{equation}

\section{Physical conditions}
\label{sec:conditions}
A configuration is admissible only if it is causal, satisfies the energy
conditions and is free of trapped surfaces. Each requirement constrains the pair
$(z,\alpha)$, and the most restrictive of them fixes the admissible region.

\subsection{Causality}

Causality requires that neither effective sound speed exceed the speed of light. From Eqs.~(\ref{eq:cs}) and~(\ref{eq:eff_explicit}), the radial sound speed is
\begin{equation}
\creff = \frac{2(1-\alpha)K\rho}{1 + 2\alpha K\rho},
\label{eq:cr_explicit}
\end{equation}
which is non-negative throughout for \(\alpha < 1\) and vanishes at the surface as required by Eq.~(\ref{eq:cond2}). Causality, \(\creff \leq 1\), gives
\begin{equation}
2(1-\alpha)K\rho \leq 1 + 2\alpha K\rho
\quad\Rightarrow\quad 2(1-2\alpha)K\rho \leq 1.
\label{eq:cr_causality}
\end{equation}
At the centre, \(\rho = \rhoc = 14B\), this imposes
\begin{equation}
2(1-2\alpha)K\rhoc \leq 1.
\label{eq:cr_central}
\end{equation}
For \(\alpha = 0\), this gives the bound on the self-interaction parameter,
\begin{equation}
2K\rhoc \leq 1 \quad\Rightarrow\quad z \leq \frac{1}{28} \simeq 0.0357,
\label{eq:z_bound}
\end{equation}
using \(K = z/B\) and \(\rhoc = 14B\). For \(\alpha > 0\), the bound is relaxed.
The tangential sound speed inherits the $\alpha$ dependence of the
anisotropy and is treated in Appendix~\ref{app:conditions}, where it is seen to
impose a joint bound upon $z$ and $\alpha$.

\subsection{Energy conditions}

The four energy conditions are written out in
Appendix~\ref{app:conditions}; their outcome is as follows. The null and weak
conditions are inherited unchanged from the uncoupled model, since
$\rhoeff + \peff = \rho + \pr$ by Eq.~(\ref{eq:inertial}) and
$\rhoeff = \rho(1 + \alpha K\rho) \geq 0$ identically. The strong condition is
weakened, as it must be wherever a negative pressure is present. The dominant
condition is the one that binds, and in the radial direction it reduces to
\begin{equation}
1 + (2\alpha-1)K\rho \geq 0.
\label{eq:dec_radial_final}
\end{equation}
which holds automatically for $\alpha \geq 1/2$ and, for
$\alpha < 1/2$, bounds $K\rhoc$ from above. With $K\rhoc = 14z$ its worst case,
$\alpha = 0$, requires $z \leq 1/14 \simeq 0.0714$. The tangential conditions
involve $\Delta$ and are imposed numerically.

Comparing Eq.~(\ref{eq:dec_radial_final}) with the causality bound of
Eq.~(\ref{eq:cr_central}) shows that the dominant condition never binds
independently. Causality requires $2(1-2\alpha)K\rhoc \leq 1$, which is
strictly stronger than the dominant condition's $(1-2\alpha)K\rhoc \leq 1$: the
former implies the latter, but not the reverse. Every configuration excluded
by the dominant energy condition is therefore already excluded by causality,
at every compactness and not only for the range of $z$ plotted in
Fig.~\ref{fig:param}.

\subsection{Absence of trapped surfaces}

The interior must remain free of trapped surfaces throughout,
\begin{equation}
\frac{2m(r)}{r} < 1 \qquad (0 \leq r \leq R).
\label{eq:no_horizon}
\end{equation}
and, $2m(r)/r$ being given explicitly in Appendix~\ref{app:conditions},
the maximum occurs in the interior and not at the surface, already at
$\alpha = 0$. The condition must therefore be imposed over the full radius,
\begin{equation}
\max_{0 \leq r \leq R} \frac{2m(r)}{r} < 1.
\label{eq:horizon_max}
\end{equation}
Setting $\alpha = 0$ and maximising gives $r_{\max}/R = \sqrt{5/6} \simeq
0.913$ and $\max_r(2m/r) = (25/24)(2M/R)$, independently of $\rhoc$ and $R$, so
that Eq.~(\ref{eq:horizon_max}) reads $2M/R < 24/25 = 0.96$ rather than
$2M/R < 1$; the vacuum term tightens it further. The surface value is therefore
necessary but not sufficient.

\ilfigwide{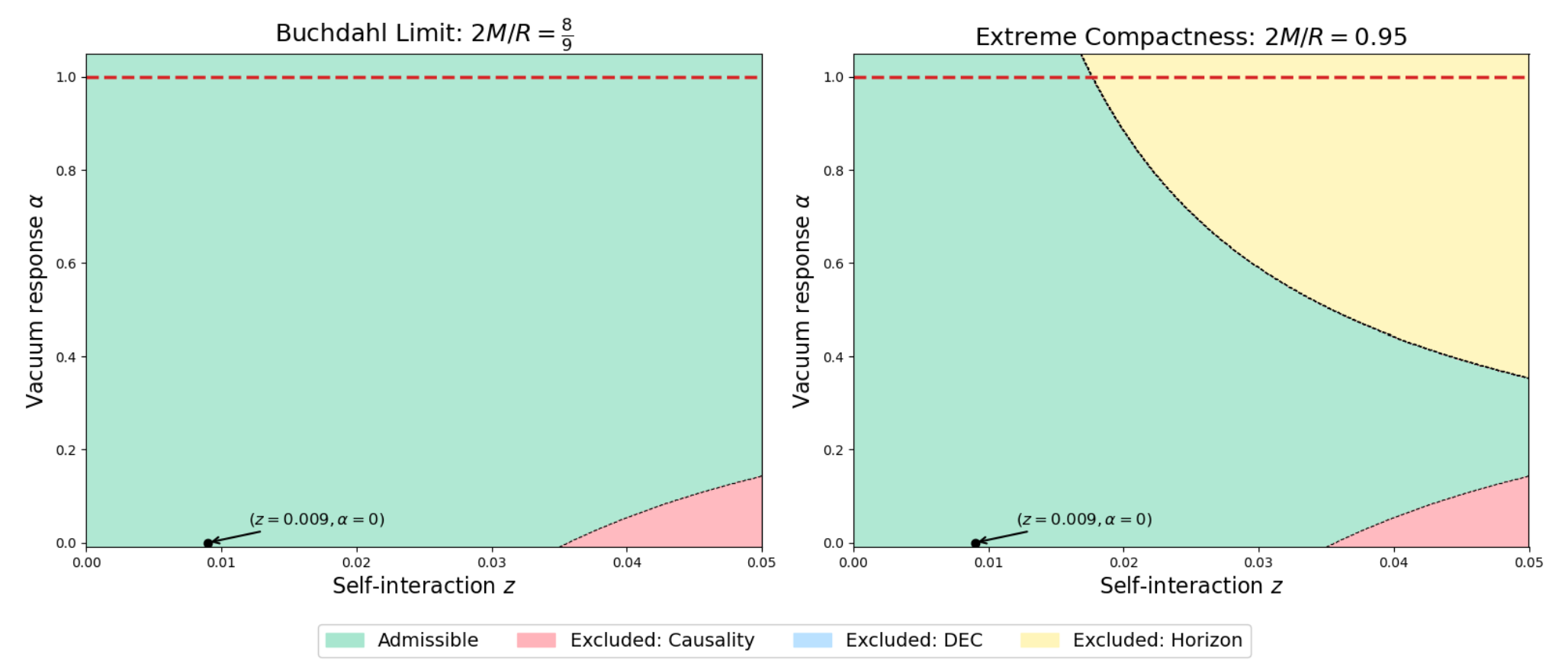}{Region of the $(z,\alpha)$ plane in which causality, the energy conditions, and the interior no-horizon condition $\max_{0 \leq r \leq R} \frac{2m(r)}{r} < 1$ are simultaneously satisfied, for the Buchdahl limit \citep{Buchdahl1959} $2M/R {|_{\alpha=0}}=8/9$ (left) and for $2M/R\ {|_{\alpha=0}}=0.95$ (right). The horizontal axis extends to $z = 0.05$, beyond the causality bound $z = 1/28 \simeq 0.0357$, to show the full exclusion region. The two panels differ only in
$\rhoc R^2$, which fixes the compactness, whereas $K\rhoc = 14z$ is independent
of it: $B$ and $R$ may be chosen separately, so no relation between the abscissa
and the compactness is implied. The dominant energy condition is not shown, since Sec.~\ref{sec:conditions} shows it to be dominated by causality throughout the plane, not only for the range plotted here.}{Abscissa: $z$, dimensionless, from zero to the causal limit. Ordinate: $\alpha$, dimensionless, over the unit interval. Shade the admissible region and hatch each excluded region separately, with a legend naming the condition responsible. Mark the parameter choice of Ref.~\citep{Moraes2021} and the limiting line $\alpha = 1$. Produced by evaluating each condition on a grid, taking the maximum of $2m(r)/r$ over the interior rather than its surface value.}

\subsection{Admissible parameter region}

The region of the \((z,\alpha)\) plane in which causality, the energy conditions, and the interior no-horizon condition are simultaneously satisfied is shown in Fig.~\ref{fig:param}. The parameter choice of Ref.~\citep{Moraes2021}, corresponding to \(\alpha = 0\) and \(z = 0.009\), lies well within the admissible region. The upper bound on $\alpha$ comes from the no-horizon condition alone.
Causality acts in the opposite sense: Eq.~(\ref{eq:cr_central}) bounds $\alpha$
from below, by $1/2 - 1/(56z)$, and only once $z > 1/28$. The dominant energy
condition, being strictly weaker than causality throughout the plane, never
binds independently. 

Figure~\ref{fig:param} presents the admissible region in the $(z,\alpha)$ plane for two values of the compactness: the Buchdahl limit ~\citep{Buchdahl1959} $2M/R |_{\alpha=0} = 8/9$ (left panel) and the extreme case $2M/R |_{\alpha=0} = 0.95$ (right panel). By the result below Eq.~(\ref{eq:horizon_max}) the latter already carries $\max_r 2m/r = 0.990$ at $\alpha = 0$. The horizontal axis extends to $z = 0.05$, beyond the causality bound $z = 1/28 \simeq 0.0357$, to show the full exclusion region. The shaded green region marks configurations that satisfy causality, the energy conditions, and the interior no-horizon condition $\max_{0 \leq r \leq R} \frac{2m(r)}{r} < 1$. The excluded regions are colour-coded: causality (red) excludes  small $\alpha$ at large $z$, and the no-horizon condition (yellow) excludes large $\alpha$. The dominant energy condition does not appear, since it is dominated by causality throughout the plane, not only for the range plotted here. The reference point of Ref.~\citep{Moraes2021}, $(z=0.009, \alpha=0)$, lies well inside the admissible region, as expected. The horizontal dashed line at $\alpha = 1$ marks the limit where the effective radial pressure vanishes identically. For the numerical evaluation, we adopt the parameters of Ref.~\citep{Moraes2021}: $B = 2 \times 10^{-80} M_{\rm Pl}^4$, $\rhoc = 14B$, and $z = 0.009$, which gives $K = z/B = 4.5 \times 10^{77} M_{\rm Pl}^{-4}$.

\ilfig{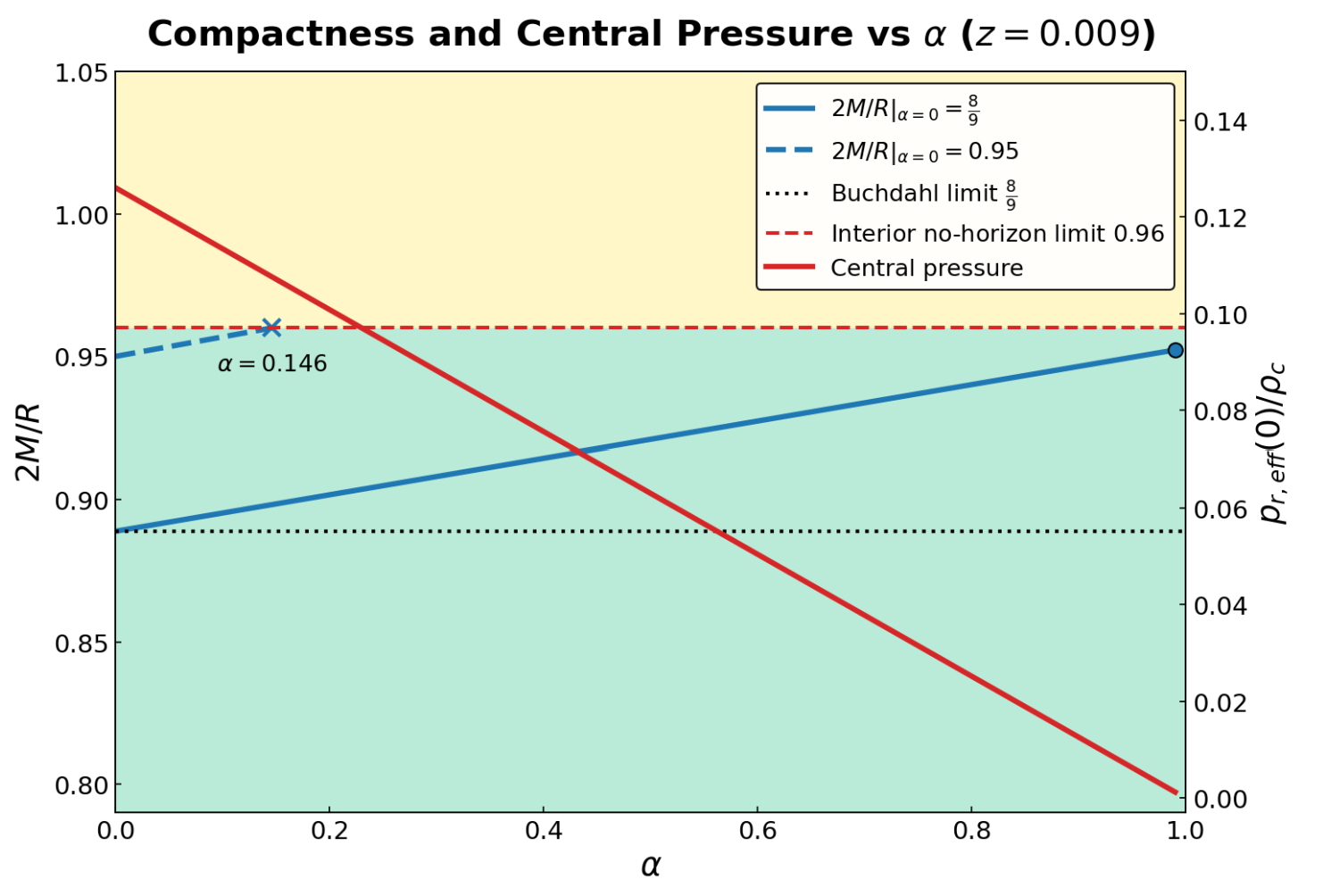}{Compactness and central effective radial pressure as functions of the vacuum response parameter, at fixed central density and radius.}{Abscissa: $\alpha$ over the unit interval. Two ordinates: $2M/R$ on the left, and $\peff(0)$ in units of the central density on the right. Produced from Eqs.~(11) and~(14). Mark the Buchdahl-type limit and the point at which the interior no-horizon condition first fails. The reader should conclude that the coupling is limited by compactness rather than by loss of pressure support.}

\section{Results}
\label{sec:results}
The first consequence is that the response does not create a pressure-sign
radius. Equation~(\ref{eq:eff_explicit}) shows that $\peff = (1-\alpha)\pr$
retains the sign of $\pr$ everywhere, so that for every admissible $\alpha$ the
configuration remains pressure supported from centre to surface. This is a
direct consequence of the surface conditions of Sec.~\ref{sec:surface}: a
response that vanishes to second order at the surface cannot overwhelm the
matter pressure anywhere, and the envelope structure obtained for a linear
response is an artifact of its inadmissibility.

The physics lies instead in the competition between two effects, and it is the
substance of this work. The vacuum response adds to the energy density, through
Eq.~(\ref{eq:mass}), whilst subtracting from the pressure, through
Eq.~(\ref{eq:eff_explicit}). The configuration therefore becomes simultaneously
heavier and less strongly supported as $\alpha$ increases, and the coupling is
bounded above not by any threshold in the pressure but by the compactness bound
of Sec.~\ref{sec:conditions}. This trade-off is set out explicitly in
Fig.~\ref{fig:compact}.

The maximum admissible response is not a single number, since it depends
upon the compactness at which the configuration is evaluated;
Fig.~\ref{fig:param} gives it as the upper edge of the admissible region. At
$z = 0.009$ the whole interval $0 \leq \alpha < 1$ remains open for
$2M/R|_{\alpha=0} = 8/9$, whereas $\alpha \lesssim 0.146$ for
$2M/R|_{\alpha=0} = 0.95$. The corresponding enhancements, from
Eq.~(\ref{eq:mass_z}), are $7.2\%$ and $1.0\%$. Figure~\ref{fig:cs} confirms
this directly.

\ilfig{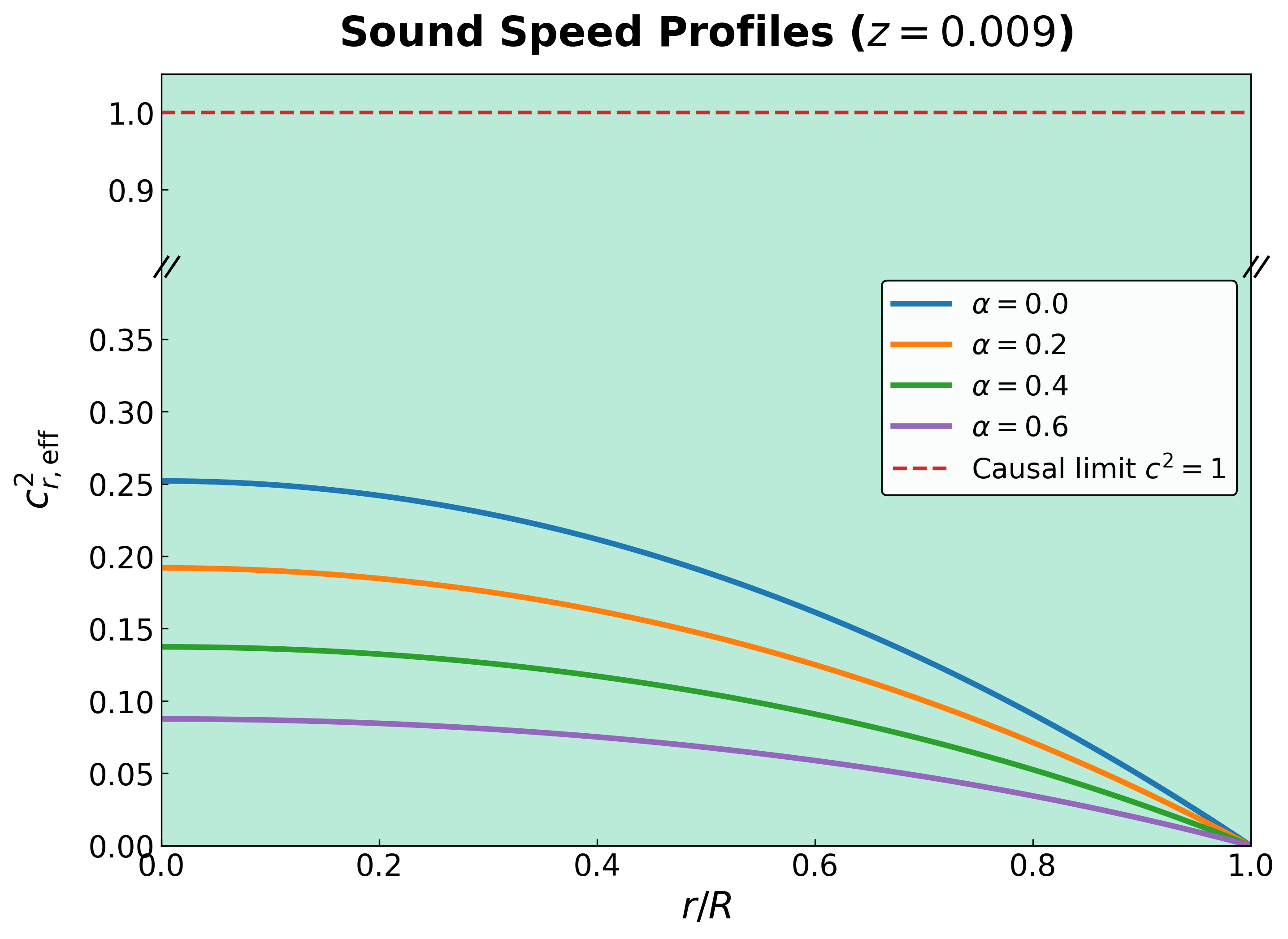}{Effective radial sound speed as a function of fractional radius, for several values of the vacuum response parameter.}{Abscissa: fractional radius over the unit interval. Ordinate: $\creff$, dimensionless. Four curves spanning $\alpha$ from zero to its maximum admissible value, with the causal limit marked. Produced from Eq.~(14). The reader should conclude that the sound speed is non-negative throughout and vanishes at the surface for every admissible $\alpha$, in contrast with the linear response.}

\ilfig{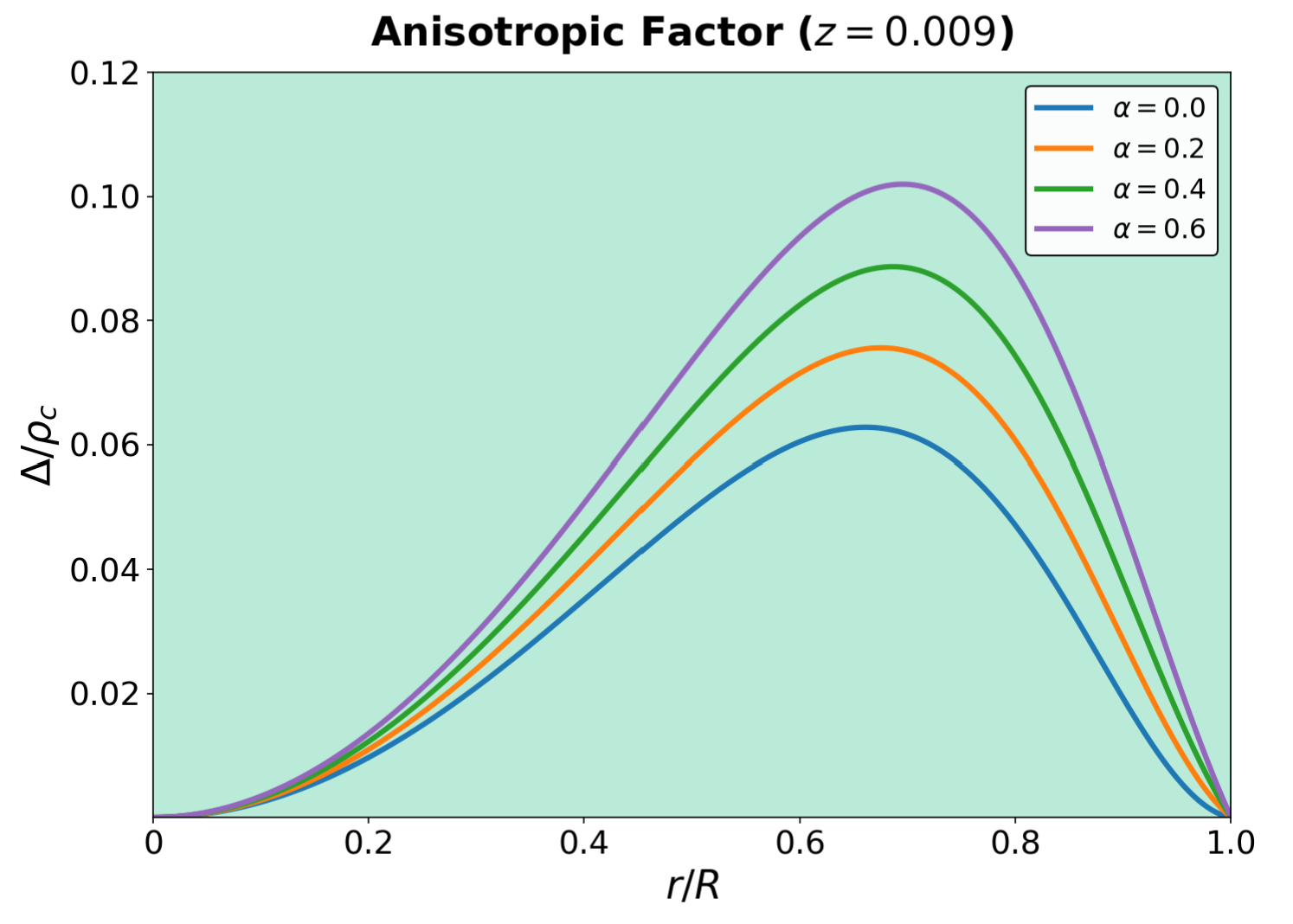}{Anisotropic factor, in units of the central density, as a function of fractional radius, for several values of the vacuum response parameter.}{Abscissa: fractional radius over the unit interval. Ordinate: $\Delta/\rhoc$, dimensionless. Four curves as before. Produced from the closed form of Eq.~(8). The reader should conclude whether the anisotropy remains positive throughout, and how its maximum shifts with $\alpha$.}

The anisotropy is determined by the conservation equation, which now carries the
exchange term of Eq.~(\ref{eq:exchange}), so its profile depends upon $\alpha$,
as Fig.~\ref{fig:delta} shows, even though $\Delta_{\rm eff} = \Delta$
formally. Regularity requires that it vanish at the centre, and the surface
conditions of Sec.~\ref{sec:surface} ensure that it vanishes at the surface
also, since $\rho$, $\pr$, $\pr'$ and $\rhoL$ all do.

\section{Discussion and conclusion}
\label{sec:conclusion}

We have asked which local vacuum responses are admissible inside an
anisotropic dark matter star, and found the stellar surface to be decisive.
Any non-negative monotone response must satisfy $f(0) = f'(0) = 0$ if the
surface is to carry a non-negative radial sound speed and no residual
anisotropy. The linear response fails the second condition, and the
pressure-sourced response $\rhoL = \alpha\pr$ is the first power law that
survives.

Adopting it, the structure equations remain in closed form. The effective
radial pressure is rescaled by $(1-\alpha)$, the energy density acquires a term
quadratic in the matter density, and the mass is enhanced by
$1 + \tfrac{4}{7}\alpha K\rhoc$. The configuration becomes heavier and less
strongly supported together, so that $\alpha$ is limited by the compactness
bound, and $\alpha \to 1$ is the limit in which radial support vanishes
identically.

This model extends the anisotropic dark matter star of
Ref.~\citep{Moraes2021} by letting the vacuum respond locally to the fluid,
subject to the surface conditions of Eqs.~(\ref{eq:cond1})
and~(\ref{eq:cond2}), which single out the pressure-sourced response of
Eq.~(\ref{eq:ansatz}) over the more obvious linear choice. That response is
itself well motivated: a repulsive, self-interacting Bose--Einstein
condensate of bosonic dark matter yields the same quadratic pressure already
adopted for Eq.~(\ref{eq:eos}), in the Thomas--Fermi
limit~\citep{BoehmerHarko2007}. Two natural extensions follow. Integrating
the structure equations with $\rhoc$ as the sole eigenvalue would deliver the
mass--radius relation and a direct comparison with observation, and a
field-theoretic derivation, for instance a scalar-field realisation of the
coupled-quintessence type~\citep{Amendola2000}, would fix $\alpha$ from a
single Lagrangian and add its own gradient contribution to the anisotropy.

A remark is in order on the stability of these configurations. Among
the many admissible choices of $\alpha$ and $z$, some will prove dynamically
stable and others will not; only the oscillation spectrum can tell them
apart, and we defer that analysis. The methods developed for anisotropic
neutron stars~\citep{Silva2015, Yazadjiev2012} offer a natural starting
point, once adapted to the background anisotropy of
Eq.~(\ref{eq:delta_explicit}) in place of the quasi-local ansatz they
employ.

\begin{acknowledgments}
I.L. thanks the Funda\c{c}\~ao para a Ci\^encia e Tecnologia (FCT), Portugal,
for the financial support to the Centre for Astrophysics and Gravitation
(CENTRA/IST/ULisboa) through grant No.~UID/PRR/00099/2025
(\doi{10.54499/UID/PRR/00099/2025}) and grant No.~UID/00099/2025
(\doi{10.54499/UID/00099/2025}).
Q.A.G. acknowledges the financial support received through the USTC Fellowship Level A–CAS-ANSO Scholarship 2025 (formerly the ANSO Scholarship for Young Talents).
\end{acknowledgments}

\appendix

\section{Reduction of the structure equations}
\label{app:structure}

This appendix collects the derivation summarised in
Sec.~\ref{sec:structure}, one physical step at a time. Throughout, $m(r)$ is
the total mass of Eq.~(\ref{eq:m_full}) and $\peff = (1-\alpha)\pr$.

\emph{Hydrostatic equilibrium.} Conservation of the total
energy-momentum tensor, $\nabla_\mu T^{\mu\nu} = 0$, gives the anisotropic
Tolman--Oppenheimer--Volkoff equation for the effective fluid,
$\peff' = -(\rhoeff+\peff)(m+4\pi r^3\peff)/[r^2(1-2m/r)] + 2\Delta/r$, where
$\Delta \equiv \pt - \pr$ and, the vacuum being isotropic, $\Delta_{\rm eff} =
\Delta$. Writing $\rhoeff+\peff = \rho+\pr$ from the inertial identity of
Eq.~(\ref{eq:inertial}), and $\peff = (1-\alpha)\pr$ from
Eq.~(\ref{eq:eff_explicit}), turns this into
\begin{equation}
(1-\alpha)\pr' = -\frac{(\rho + \pr)[m + 4\pi r^3 (1-\alpha)\pr]}{r^2(1 - 2m/r)}
+ \frac{2\Delta}{r},
\label{eq:tov_alpha}
\end{equation}
whose division by $(1-\alpha)$ is Eq.~(\ref{eq:tov_final}) of the
body.

\emph{The metric potential.} The $(rr)$ Einstein equation gives
$\nu' = 2(m+4\pi r^3\peff)/[r^2(1-2m/r)]$, which with $\peff = (1-\alpha)\pr$
reads
\begin{equation}
\nu' = \frac{2[m + 4\pi r^3 (1-\alpha)\pr]}{r^2(1 - 2m/r)}.
\label{eq:nu_prime_alpha}
\end{equation}
Matching to the exterior Schwarzschild metric fixes the surface value,
$e^{\nu(R)} = 1 - 2M/R$, and integrating outward from the surface gives
\begin{equation}
\nu(r) = \ln\left(1 - \frac{2M}{R}\right)
+ 2\int_R^r \frac{m(x) + 4\pi x^3 (1-\alpha)\pr(x)}
{x^2(1 - 2m(x)/x)}\,dx.
\label{eq:nu_integral}
\end{equation}

\emph{Junction conditions.} The Israel--Darmois
conditions~\citep{Israel1966} require continuity of the induced metric and of
the extrinsic curvature across $r=R$, so that a matching without a surface
layer needs $\peff(R^-) = \peff(R^+)$. Since $\peff(R^-) = (1-\alpha)\pr(R) =
0$ and the exterior is vacuum with $\peff(R^+) = 0$, this holds identically.
Moreover $\rhoL(R) = \alpha\pr(R) = 0$ and $\rho(R) = 0$, so the energy
density is continuous as well, and no thin shell is required.

\emph{The anisotropic factor.} Inverting Eq.~(\ref{eq:tov_alpha}) for
$\Delta$, and substituting the explicit forms of $\rho$, $\pr$ and $m$ from
Eqs.~(\ref{eq:eos}) and~(\ref{eq:m_full}), yields the closed form
\begin{widetext}
\begin{equation}
\begin{split}
\Delta(r) = \frac{r}{2}\Bigg\{
&-4(1-\alpha)K\rhoc^2 \frac{r}{R^2}\left(1 - \frac{r^2}{R^2}\right) \\
&+ \frac{[\rho(r)+\pr(r)]\left[m(r) + 4\pi r^3(1-\alpha)K\rhoc^2\left(1 - \frac{r^2}{R^2}\right)^2\right]}
{r^2\left(1 - 2m(r)/r\right)}
\Bigg\},
\end{split}
\label{eq:delta_explicit}
\end{equation}
\end{widetext}
with $m(r)$ from Eq.~(\ref{eq:m_full}). Its boundary values are quoted
in Sec.~\ref{sec:structure}.

\section{The physical conditions in full}
\label{app:conditions}

This appendix carries the conditions summarised in
Sec.~\ref{sec:conditions}, condition by condition.

\emph{The tangential sound speed.} Differentiating
Eq.~(\ref{eq:pt_eff}) with respect to $\rhoeff$ gives
\begin{equation}
\cteff =  \creff + \frac{d\Delta}{d\rhoeff},
\label{eq:ct_chain}
\end{equation}
where $\creff$ is given by Eq.~(\ref{eq:cr_explicit}) and
$d\Delta/d\rhoeff$ is evaluated as the quotient of radial derivatives,
$(d\Delta/dr)/(d\rhoeff/dr)$, which is regular for $0 < r \leq R$. Causality
requires $0 \leq \cteff \leq 1$ for $0 \leq r \leq R$, and this condition,
through its dependence upon $\Delta$ from Eq.~(\ref{eq:delta_explicit}),
imposes a joint bound upon $z$ and $\alpha$.

\emph{Null energy condition.} The condition $\rhoeff + p_{i,\rm eff}
\geq 0$ must hold for each principal pressure. Radially, $\rhoeff+\peff =
\rho+\pr \geq 0$, which is satisfied identically by Eq.~(\ref{eq:eos}).
Tangentially, $\rhoeff + \pteff = \rho + \pt \geq 0$, or, using
Eq.~(\ref{eq:pt_eff}),
\begin{equation}
\rho + \pr + \Delta \geq 0,
\label{eq:nec_tangential_explicit}
\end{equation}
which is not guaranteed by the equation of state and is imposed
numerically.

\emph{Weak energy condition.} This requires $\rhoeff \geq 0$ in
addition to the null condition. Since $\rhoeff = \rho(1+\alpha K\rho) \geq
0$, the weak condition reduces to the null conditions.

\emph{Strong energy condition.} This requires $\rhoeff + \peff +
2\pteff \geq 0$. Substituting the effective quantities and simplifying
gives
\begin{equation}
\rho + (3 - 2\alpha)\pr + 2\Delta \geq 0.
\label{eq:sec_final}
\end{equation}
The strong condition is weakened relative to the uncoupled case by the
presence of the negative vacuum pressure.

\emph{Dominant energy condition.} This requires $\rhoeff \geq
|\peff|$ and $\rhoeff \geq |\pteff|$. Radially, $\rho(1+\alpha K\rho) \geq
(1-\alpha)K\rho^2$, which after dividing by $\rho > 0$ becomes
\begin{equation}
1 + \alpha K\rho \geq (1-\alpha)K\rho,
\label{eq:dec_radial_simp}
\end{equation}
which recovers Eq.~(\ref{eq:dec_radial_final}) of the body.
Tangentially,
\begin{equation}
\rho(1+\alpha K\rho) \geq |(1-\alpha)K\rho^2 + \Delta|,
\label{eq:dec_tangential}
\end{equation}
which is likewise imposed numerically.

\emph{Absence of trapped surfaces.} Using Eq.~(\ref{eq:m_full}),
Eq.~(\ref{eq:no_horizon}) becomes
\begin{widetext}
\begin{equation}
\frac{2m(r)}{r} = 8\pi\rhoc r^2\left(\frac{1}{3} - \frac{r^2}{5R^2}\right)
+ 8\pi\alpha K\rhoc^2 r^2\left(\frac{1}{3} - \frac{2r^2}{5R^2} + \frac{r^4}{7R^4}\right) < 1.
\label{eq:horizon_explicit}
\end{equation}
\end{widetext}
whose maximum over the interior is the quantity constrained by
Eq.~(\ref{eq:horizon_max}). The surface staticity condition $2M/R < 1$ is
necessary but not sufficient, since the interior maximum may exceed the
surface value.

\bibliographystyle{apsrev4-2}
\bibliography{DraftQL}

\end{document}